\documentclass[superscriptaddress,twocolumn,prb]{revtex4}

\usepackage{graphicx}
\usepackage{verbatim}
\usepackage{comment}
\usepackage{amsmath,amssymb,relsize}

\def\ket#1{\mathinner{|{#1}\rangle}}

\makeatletter
\def\Ddots{\mathinner{\mkern1mu\raise\p@
\vbox{\kern7\p@\hbox{.}}\mkern2mu
\raise4\p@\hbox{.}\mkern2mu\raise7\p@\hbox{.}\mkern1mu}}
\makeatother
\usepackage{amsmath}
\usepackage{amsthm}
\usepackage{amssymb}
\usepackage{rotating}
\usepackage{epstopdf}
\usepackage{bbold}
\usepackage{times}
\usepackage[none]{hyphenat}
\usepackage{threeparttable}

\usepackage{color}
\usepackage{mathrsfs}

\usepackage[normalem]{ulem}

\begin{document}

\author{Giuseppe Pica$^\ast$}
\affiliation{SUPA, School of Physics and Astronomy, University of St Andrews, KY16 9SS, United Kingdom} 
\author{Brendon W. Lovett}
\affiliation{SUPA, School of Physics and Astronomy, University of St Andrews, KY16 9SS, United Kingdom}

\title{Theory of single and two-qubit operations with donor-bound electron spins in germanium}

\begin{abstract}
The possibility of quantum computing with spins in germanium nanoscale transistors has recently attracted interest since it promises highly tuneable qubits that have encouraging coherence times. We here present the first complete theory of the orbital states of Ge donor electrons, and use it to show that Ge could have significant advantages over silicon in the implementation of a donor-based quantum processor architecture. We show that the stronger spin-orbit interaction and the larger electron donor wave functions for Ge donors allow for greater tuning of the single qubit energy than for those in Si crystals, thus enabling a large speedup of selective (local) quantum gates. Further, exchange coupling between neighboring donor qubits is shown to be much larger in Ge than in Si, and we show that this greatly relaxes the precision in donor placement needed for robust two-qubit gates. To do this we compare two statistical distributions for Ge:P and Si:P pair couplings, corresponding to realistic donor implantation misplacement, and find that the spin couplings in Ge:P have a $33\%$ chance of being within an order of magnitude of the largest coupling, compared with only $10\%$ for the Si:P donors. This allows fast, parallel and robust architectures for quantum computing with donors in Ge. \\ \let\thefootnote\relax\footnotetext{$^\ast$Current address: Center for Neuroscience and Cognitive Systems, Italian Institute of Technology, Corso Bettini 31 - 38068 Rovereto, Italy} 
\end{abstract}

\maketitle

\section{Introduction}
The recent progress of quantum computing with electron and nuclear spins in doped Si~\cite{RevModPhys.85.961}, with demonstrated record coherence in the solid state~\cite{tyryshkin12,saeedi13}, has been astonishing. In spite of this, the intrinsically weak spin-orbit interaction of the conduction electrons with the host Si nuclei~\cite{PhysRevB.71.075315} and the relatively tight wave functions of the donor electrons~\cite{PhysRevB.90.195204} can seriously restrict the extent to which the qubit states can be selectively controlled. 
Moreover, coupling two donor spins in Si requires the ions to be implanted very closely~\cite{kane98}, and with high precision~\cite{oscilla2}. 

While retaining the compatibility with integrated circuits, Ge spins can be efficiently controlled electrically, thanks to the stronger spin-orbit interaction: theoretical predictions of high $g$-factor tunability of conduction band electrons in Ge-based heterostructures~\cite{PhysRevB.68.195306} have been confirmed by very recent experiments~\cite{2016arXiv160308783G}. Despite this decreased spin isolation, recent electron spin resonance (ESR) measurements of isotopically-purified-Ge donors have shown promising coherence times $T_{2}$$\sim$ms~\cite{PhysRevLett.115.247601}. Furthermore, Ge has been considered as a candidate host for transistor processing in the quantum regime~\cite{PhysRevA.62.012306}, and the technical development in device fabrication parallels the more popular Si electronics~\cite{Scappucci2011,gemosfets}. We are thus motivated to develop a complete theory of electron states of Ge donors, with the aim of predicting whether and by how much this new semiconductor platform could improve and facilitate the control of single and two donor qubits.

Extensive ESR experiments with Ge:P, Ge:As and Ge:Bi donors at low temperature were already performed by Wilson~\cite{wilson} in the '60s. However, the theoretical understanding of Ge donors is not well developed: the most modern approaches to describing the electron orbital states date back to the '70s~\cite{PhysRev.184.713,PhysRevB.1.4673,altarelliC0}, with a satisfactory theory still missing to date. Nonetheless, a detailed picture is crucial for deciding whether Ge-donor spins could make promising qubits, and for anticipating the most desirable regimes for spin manipulation in such devices. In this paper, we provide for the first time a thorough description of shallow Ge-donors from group V, based on an improved multi-valley effective mass theory (MV-EMT) that includes the strong anisotropy of the conduction band and approximates phenomenologically the interaction between the donor electron and the nucleus~\cite{ningsah,PhysRevB.90.195204}. Unlike in most of the traditional EMT papers, here the full plane-wave expansion of the Bloch states of the pure Ge crystal is included: this provides a crucial step for developing a unified picture across different donor chemical species.

We then model DC gate control of the spin resonance frequencies~\cite{kane98,PhysRevLett.97.176404,Lauchte1500022}, by calculating the susceptibilities of both the hyperfine and the Zeeman splittings for all group V donors. Close agreement with the contemporary experimental measurements of Ge:As by Sigillito {\it et al.} in Ref.~\onlinecite{sigillito16} validates our estimates of the electrical tuning of donor spin splitting. We predict that single qubit gates with Ge:P donors could be performed with nanosecond timescales under realistic assumptions, a huge improvement to the Si framework. This effect was partially anticipated by earlier \emph{ab initio} treatments in Ref.~\onlinecite{PhysRevB.80.155301}, but we improve those predictions with quantitative matching of experiments, and extend them to all donors with a more clear understanding of the underlining physics. 

We further move on to calculate for the first time the exchange couplings $J$ between adjacent Ge-donor spins, and quantify how their larger wave functions allow the stringent pitch requirements of Si-donors' placement to be relaxed by a factor of three. Remarkably, we show that, in the optimal geometric configuration, $J$ couplings between pairs of Ge-donors will depend less than those in Si on 3D implantation misplacements. Uniform control of large, strongly coupled qubit clusters thus promises to be significantly easier in Ge, which we predict to be more robust in the face of the problems of spin coupling intrinsic to multi-valley semiconductors~\cite{oscilla2,PhysRevB.89.235306,PhysRevB.91.235318}.

\section{Multi-valley effective mass theory of donors in Ge}\label{mvemt}
The lowest conduction band in Ge has four equivalent minima (valleys) $\textbf{k}_{0\mu}$, each located at the edge of the Brillouin zone along one of the equivalent crystallographic $\langle 111 \rangle$ directions in $k$ space, the so-called $L$ points~\cite{yucardona}. We will choose $\{\textbf{k}_{0\mu}\}=\frac{\pi}{a_{\rm Ge}}\{(111),(\bar{1}11),(1\bar{1}1),(11\bar{1})\}$, where $a_{\rm Ge}=0.566$ nm is the crystal lattice constant~\cite{yucardona}. The ground state of a conduction electron in the undoped Ge crystal is thus four-fold degenerate as regards the orbital degree of freedom. However, the implantation of a substitutional atom from group V partially breaks the crystal symmetry, as the Hamiltonian $H^{0}$ of the donor electron includes an extra term $U(\textbf{r})$, that describes its interaction with the impurity ion. Effective mass theory (EMT) maintains that the donor quantum states can still be expanded in terms of packets of Bloch functions whose $\textbf{k}$-vectors concentrate around each minimum, since the wave function is known to spread across tens of lattice constants in space. While this approximation fails to  accurately describe the behavior of the donor electron wave function within the crystal cell occupied by the binding substitutional impurity (the {\it central cell}), it does not impede a reliable picture outside of it, if one averages the short-range Hamiltonian with a pseudopotential~\cite{ningsah,PhysRevB.90.195204}. Importantly, both single and two-qubit manipulations of donor spins rely on the long-range component of the donor wave function, since it is this part that is affected by external control fields and/or overlap with neighboring donor states.

Over the years, EMT has been improved from its original single-valley formulation~\cite{kl1,kl2} to a multi-valley framework~\cite{ningsah,Pantelides1974,shindonara,Debernardi2006,hui,PhysRevB.90.195204,PhysRevB.91.235318}, which aims at describing how all valleys are admixed in the electron eigenstates as a result of the donor-dependent impurity potential $U(\textbf{r})$. The most important consequences for the electron ground state are that it is non-degenerate, and its orbital energy and hyperfine coupling to the ion differ significantly from a single-valley state~\cite{ramdas,wilson}. The theoretical progress in the understanding of Si-donors has not been paralleled by adequate investigation of Ge-donors, where the electron wave functions are more spread out because of both stronger dielectric screening from the crystal and smaller average effective mass of the lowest conduction band. This is evident if one compares the rescaled effective Rydberg energies and transverse Bohr radii of the orbital states in Si ($a^\ast_{Si}=\hbar^2\epsilon_{Si}/(m^\ast_{\perp,Si}e^2)=3.157$ nm, $Ry^\ast=m^\ast_{\perp,Si}e^4/2\hbar^2\epsilon^2_{Si}=19.98$ meV) and Ge ($a^\ast_{\rm Ge}=10.244$ nm, $Ry^\ast=4.45$ meV).

We now present the most important steps that lead to the final multi-valley EMT equations that we are going to use, starting from the exact Schr\"odinger equation for a donor electron in bulk Ge: 
\begin{equation}\label{eqbase}
H^{0} \Psi(\textbf{r})= \left[-\frac{\hbar^{2}}{2 m_{0}}\nabla^{2} + V^{0}(\textbf{r}) + U(\textbf{r}) \right]\Psi(\textbf{r})= \epsilon \Psi(\textbf{r}),
\end{equation}   
where $\Psi(\textbf{r})$ is the wave function of the donor electron, $m_{0}$ is its rest mass, $V^{0}(\textbf{r})$ is the periodic potential of the undoped Ge crystal, and $\epsilon$ labels any energy eigenvalue.

EMT dictates that $\Psi(\textbf{r})$ be expanded as a sum of the Bloch functions of the conduction electrons in undoped Ge $\phi_{0}(\textbf{k},\textbf{r})\equiv u_{0}(\textbf{k},\textbf{r}) e^{i \textbf{k}\cdot \textbf{r}}$, each weighted with envelopes $\tilde{F}_{\mu}$ that decay rapidly as their argument gets farther from each of the four valleys $\textbf{k}_{0\mu}$ ~\cite{kl1}:
\begin{align}\nonumber
\Psi(\textbf{r}) &\equiv \sum_{\mu}\alpha_{\mu} \xi_{\mu}(\textbf{r}) \\
\label{waveexpansion} &= \sum_{\mu}\alpha_{\mu}\frac{1}{(2 \pi)^{3}} \int \tilde{F}_{\mu}(\textbf{k}_{\mu}+\textbf{k}_{0\mu})\phi_{0}(\textbf{k}_{\mu}+\textbf{k}_{0\mu},\textbf{r}) d\textbf{k}_{\mu},
\end{align}  
where we have conveniently defined $\xi_{\mu}$ as the total contribution of each envelope centered at $\textbf{k}_{0\mu}$, and $\alpha_{\mu}$ are valley coefficients that can be predicted by symmetry arguments alone~\cite{wilson}. In fact, the four-fold $1s$ valley manifold can be grouped into a singlet $A_{1}$ and a triplet $T_{2}$:
\begin{equation}\label{groupstuff}
\begin{array}{l}
\alpha_{i}(A_{1})=\frac{1}{2}(1,1,1,1) \\
\\
\alpha_{i}(T_{2})=\left\{\begin{array}{l}\frac{1}{\sqrt{2}}(1,-1,0,0)\\
\frac{1}{2}(1,1,-1,-1)\\
\frac{1}{\sqrt{2}}(0,0,1,-1).\\ 
\end{array}\right. 
\end{array}
\end{equation}
A further step of EMT is to encode the Hamiltonian of the pure crystal in an effective mass tensor specific to the lowest conduction band: this contributes an anisotropic kinetic energy operator $\textbf{p}\cdot\textbf{A}_{\mu}\cdot\textbf{p}$ for each $\hat{\mu}$ valley. Here, $\textbf{p}$ is the momentum operator of the donor electron, and $\textbf{A}_{\mu}$ is a diagonal tensor in a basis with one vector parallel to $\hat{\mu}$ and two vectors orthogonal to it (e.g., if $\hat{\mu}\parallel [111]$, one could choose the basis $\{[1\bar{2}1],[10\bar{1}],[111]\}$), with entries $A_{\mu}^{11}=A_{\mu}^{22}=1/(2 m^{\ast}_{\perp})$ and $A_{\mu}^{33}=1/(2 m^{\ast}_{\parallel})$, where $m^{\ast}_{\perp}=0.0815$ $m_{0}$ and $m^{\ast}_{\parallel}=1.59$ $m_{0}$.

We are thus ready, following Ref.~\onlinecite{shindonara}, to take the expectation values of both sides of Eq.~(\ref{eqbase}) for the wave function in Eq.~(\ref{waveexpansion}): after performing the $\textbf{k}$ integrations and introducing the Fourier transformed envelopes $F_{\mu}(\textbf{r})$, this gives 
\begin{gather}\nonumber
\int d\textbf{r} \sum_{\textbf{G}}\sum_\mu\alpha^{\ast}_{\mu}F^{\ast}_{\mu}(\textbf{r}) \times [\alpha_{\mu} (\textbf{p}\cdot \textbf{A}_{\mu}\cdot \textbf{p}-\epsilon) F_{\mu}(\textbf{r}) +\\ 
\label{definitiva}\sum_\nu \alpha_{\nu} e^{-i(\textbf{k}_{0\mu}-\textbf{k}_{0\nu})\cdot \textbf{r}} C_{\textbf{G}}(\textbf{k}_{0\nu},\textbf{k}_{0\mu}) e^{i \textbf{G}\cdot \textbf{r}} U(\textbf{r}) F_{\nu}(\textbf{r})] =0  ,
\end{gather} 
where the labels $\mu$ and $\nu$ each run across the four valley minima, and $\sum_{\textbf{G}}C_{\textbf{G}}(\textbf{k},\textbf{k}') e^{i \textbf{G}\cdot \textbf{r}}\equiv u_{0}^{\ast}(\textbf{k},\textbf{r})u_{0}(\textbf{k}',\textbf{r})$ is the expansion of the lattice-periodic portion of the Bloch functions in terms of the vectors $\{\textbf{G}\}$ of the Ge reciprocal lattice. We have also assumed that $C_{\textbf{G}}(\textbf{k}_{0\mu}+\textbf{k}_{\mu},\textbf{k}_{0\nu}+\textbf{k}_{\nu})\approx C_{\textbf{G}}(\textbf{k}_{0\mu},\textbf{k}_{0\nu})$, as it is found that these weights vary weakly for small displacements of their arguments.  

We combine this multi-valley EMT with a pseudopotential method to describe the impurity potential $U(\textbf{r})$, an approach that has been applied successfully to understand some properties of donor electron states in Si~\cite{PhysRevB.90.195204}. The use of a smooth pseudopotential in place of the true impurity potential implies that the eigenstates of the corresponding Hamiltonian are not accurate on the scale of the central cell~\cite{ningsah}, where strong oscillations impede the semi-analytic description we are aiming at. However, the pseudo-wave function averages the short-ranged behaviour and matches it consistently with the more accurate picture farther from the nucleus. The past decades have seen a plethora of functional forms proposed for the impurity pseudopotential, whereby few adjustable parameters were tuned until the low-energy donor states matched one or more experimental binding energies. However, we recently pointed out that the variational solutions of the corresponding pseudo-Hamiltonians are not fundamentally granted to provide sensible descriptions of the donor \emph{eigenstates} of the problem in Eq.~(\ref{eqbase}), although they can give reliable upper bounds to its exact \emph{eigenvalues}\cite{PhysRevB.89.235306,PhysRevB.90.195204}. In order to assess donor architectures for quantum computing, enough accuracy is needed in the description of some key features of the donor wave function, while at the same time keeping the theoretical framework light enough that those features can be flexibly evaluated within different realistic environments. Thus we suggested that, within an analytical MV-EMT/pseudopotential framework, the tunable parameters in $U(\textbf{r})$ should also fit the hyperfine coupling to the impurity ion, in order to constrain the average short-ranged behaviour of the pseudo-state, and, crucially, to improve the description of the eigenstates. We require, then, a phenomenological pseudopotential which is the simplest possible that allows for a unified and accurate framework across all donors. We find that the following short-ranged pseudopotential with spherical symmetry of the form:
\begin{equation}\label{ning}
U(\textbf{r})=-\frac{e^{2}}{\epsilon_{\rm Ge}|\textbf{r}|}(1-\text{e} ^{- b |\textbf{r}|}+ B |\textbf{r}| \text{e} ^{- b |\textbf{r}|}) \equiv -\frac{e^{2}}{\epsilon_{\rm Ge}|\textbf{r}|}+U_{cc}(\textbf{r}),
\end{equation}
meets our needs. Here, $\epsilon_{\rm Ge}=16.2$ is the static dielectric constant for Ge, $e$ is the elementary charge, and $b$ and $B$ are two donor-dependent parameters that will be fit, once for all, to the experimental binding energies and hyperfine coupling.  

Unlike our previous work on Si donors in Ref.~\onlinecite{PhysRevB.90.195204}, we find that a consistent picture requires an important further improvement, as we now explain. Most of the previous EMT approaches go on to neglect the $\textbf{G}\neq 0$ terms of the plane-wave expansion of the Bloch states $u_0(\textbf{k},\textbf{r})$~\cite{shindonara}, as they mostly multiply higher Fourier components of the impurity potential in Eq.~(\ref{definitiva}). However, it is not always true that these terms are less important than the $\textbf{G}= 0$ terms, and this approximation has been criticized~\cite{resta,PhysRevB.91.235318}. The assumption is particularly questionable for Ge donors, since the most important Bloch states $u_0(\textbf{k}\sim\textbf{k}_{0\mu},\textbf{r})$ in Eq.~(\ref{waveexpansion}) in Ge are even farther from $\textbf{k}=0$ than in Si. Thus, Umklapp scattering terms from the periodic lattice potential are more important, as it is easier for reciprocal lattice vectors $\textbf{G}$ to `resonantly' match differences $\textbf{k}-\textbf{k}'$ between two momenta belonging to separate Brillouin zones. The inter-valley weights $C_{\textbf{G}}(\textbf{k}_{0\nu},\textbf{k}_{0\mu})$ ($\mu\neq\nu$) with $\textbf{G}\neq 0$ are thus expected to be more significant, as direct pseudopotential calculations show~\cite{altarelliC0,cohenbergstresser}.

 
Our solution is to include the full Bloch structure of the donor states: the list of the relevant weights $C_{\textbf{G}}(\textbf{k}_{0\mu},\textbf{k}_{0\nu})$ has been evaluated in Ref.~\onlinecite{altarelliC0} after a pseudopotential calculation of the band structure of the undoped Ge crystal~\cite{cohenbergstresser}. We report the relevant results in Table~\ref{tab:intervalleycoupling} for convenience. Specifically, we neglect those terms that either have vanishing weights ($|C_{\textbf{G}}(\textbf{k}_{0\mu},\textbf{k}_{0\nu})|\lesssim 0.1$) or that sample very high Fourier components $\tilde{U}(\textbf{k}_{0\mu} - \textbf{k}_{0\nu}-\textbf{G}\gtrsim 4 k_0 )$ and are thus negligible. We verify \emph{a posteriori} that the numerical results obtained here are not influenced significantly by these approximations.
\begin{table}
\setlength{\belowcaptionskip}{5pt}
\begin{minipage}[l]{.2\textwidth}
  \begin{tabular}{l@{\hspace{4pt}} *{2}{c}}
\hline 
\textbf{G} shell & $C_{\textbf{G}}(\textbf{k}_{0\mu},\textbf{k}_{0\nu}), \mu\neq \nu$   \\
\hline
$\langle000\rangle$ & 0.42 \\
\hline
$\langle111\rangle$ & 0.35  \\
\hline
$\langle200\rangle$ & 0.0 \\
\hline
$\langle220\rangle$ & 0.95 \\
\hline
$\langle311\rangle$ & 0.39 \\
\end{tabular}
\end{minipage}\hspace{6mm}
\begin{minipage}[l]{.23\textwidth}
\vspace{-4mm}\caption{Intervalley coupling weights $C_{\textbf{G}}(\textbf{k}_{0\mu},\textbf{k}_{0\nu})$ due to the crystal periodic parts of the Bloch functions pinned to any two different Ge valleys $\mu\neq\nu$, scanned at different $\textbf{G}$ shells. Adapted from Ref.~\onlinecite{altarelliC0}.}
\label{tab:intervalleycoupling}
\end{minipage}
\end{table}

\subsection{Bulk $1s$ donor states}
We solve variationally for the eigenvalue multi-valley EMT equation~(\ref{definitiva}): once two values for $b$ and $B$ are chosen, trial envelopes $F_{\mu}^{0}$ that suitably reproduce the anisotropic symmetry of each valley are optimized to minimize the expectation value of the energy over the corresponding state. If we consider for example the $[111]$ valley, and orient the spatial $\{x,y,z\}$ Cartesian axes along $\{[1\bar{2}1],[10\bar{1}],[111]\}$ respectively, our \emph{ansatz} is written as 
\begin{equation}\label{newtrial}
F^{0}_{111}(x,y,z)= N_{0} \left[\text{e}^{-\sqrt{\frac{x^{2}+y^{2}}{a^{2}_{s}}+\frac{z^{2}}{b^{2}_{s}}}} + \beta \hspace{1mm} \text{e}^{-\sqrt{\frac{x^{2}+y^{2}}{a^{2}_{l}}+\frac{z^{2}}{b^{2}_{l}}}}\right], 
\end{equation}    
Here, $N_{0}$ is a suitable normalization that implies $\int_{\text{all space}} d\textbf{r}$ $|F_{111}^0(\textbf{r})|^2=1$, while the other coefficients are five variational parameters. We use the superscript `0' to distinguish this function from one we use shortly to model an applied electric field. Agreement with experimental binding energies and hyperfine coupling of the ground state for each donor species, as reported in Tables ~\ref{tab:fitting} and ~\ref{tab:hyperfine}, sets the best fit values of $b$ and $B$ for each donor. Even though our pseudo wave function is not expected to provide a trustworthy account of the electron state in the central cell, a rough extrapolation to the value of $\Psi(0)$ that is needed to compute the contact hyperfine coupling between the electron and the nucleus can be attained by including a multiplicative bunching factor of Bloch functions at the lattice sites, which estimates the local value of the true wave function at the impurity site~\cite{kohnbook}. $\eta_{\rm Ge}\sim9.5 \eta_{Si}$ has been extracted from comparative measurements in Ref.~\onlinecite{wilson}, and we use the latest value $\eta_{Si}\approx159$ as calculated with tight binding methods in Ref.~\onlinecite{assali}.

\begin{table}[t]
\begin{center}
  \begin{tabular}{l@{\hspace{10pt}} *{5}{c}}
\hline 
\textbf{Donor} & b (nm$^{-1}$) & B (nm$^{-1}$) & $\epsilon_{A_{1}}^{exp}$ (meV)~\cite{wilson}& $\epsilon_{A_{1}}^{th}$  (meV)\\ 
\hline
P & 4.4 & 25.9 & -12.89 & -12.96 \\
\hline
As & 5.8 & 43.9 & -14.17 & -13.52 \\
\hline
Bi & 8.8 & 69.3 & -12.75 & -13.26 \\
\hline
\end{tabular}
\caption{Fitting of the pseudopotential: the parameters $b$ and $B$ in Eq.~\ref{ning} are chosen for each donor so that the theoretical ground energy $\epsilon_{A_{1}}^{th}$ and hyperfine coupling (proportional to $|\Psi(0)|^2_{th}$) match with their experimental values $\epsilon_{A_{1}}^{exp}$ and $A_{0}$.}
\label{tab:fitting}
\end{center}
\end{table}

We list the optimal parameters of each ground donor wave function, as defined in Eqs.~(\ref{waveexpansion}) and~(\ref{newtrial}), in Table~\ref{tab:parameters}. Let us remark that different subsets of each valley manifold (e.g. the orbital singlet $A_1$ and triplet $T_2$), and of course different orbital states, will require different values for the optimal variational parameters.\\

\begin{table}[t]
  \begin{tabular}{l@{\hspace{10pt}} *{4}{c}} 
\hline
\textbf{Donor} & A$_{0}$ (MHz) ~\cite{wilson} & $|\Psi(0)|^{2}_{exp} (cm^{-3})$ ~\cite{wilson} & $|\Psi(0)|^{2}_{th} (cm^{-3})$ \\
\hline
P & 45.9  & $0.17\times10^{24}$ & 0.19 $\times10^{24}$ \\
\hline
As &  78.0 & $0.69\times10^{24}$ & 0.64 $\times10^{24}$ \\
\hline
Bi & 229 & $2.15\times10^{24}$ & 2.09$\times10^{24}$ \\
\hline
\end{tabular}
\caption{Theoretical values are calculated using $|\Psi(0)|_{th}^{2}=4$ $ \eta_{\rm Ge} |F^{0}(0)|^{2}$, where $\eta_{\rm Ge}=|u_{0}(\textbf{k}_{0},0)|^{2}/\langle |u_{0}(\textbf{k}_{0},\textbf{r})|^{2}\rangle_{\text{unit cell}} =9.5$ $ \eta_{\rm Si}$ as measured in Ref.~\onlinecite{wilson}, and $\eta_{\rm Si}=159.4$ is taken from Ref.~\onlinecite{assali}. $F^{0}$ is any of the envelopes $F^{0}_\mu(0)$, for example that in Eq.~\ref{newtrial}, since all of them are the same at ${\bf r}=0$ } 
\label{tab:hyperfine}
\end{table}
\begin{table}[h]
\begin{center}
  \begin{tabular}{l@{\hspace{4mm}} *{6}{c}}
\hline 
\textbf{Donor} & $\bar{a}_{s}$ (nm) & $\bar{b}_{s}$ (nm) & $\bar{\beta}$ & $\bar{a}_{l}$ (nm) & $\bar{b}_{l}$ (nm)  \\
\hline
P & 0.666  & 0.358  & 1.29 & 4.69 & 1.41 \\
\hline
As & 0.348 & 0.123 & 0.597 & 4.14 & 1.24  \\
\hline
Bi & 0.184 & 0.064 & 0.318 & 3.76 & 1.11  \\
\hline
\end{tabular}
\caption{Wave function parameters for the variationally optimized donor ground state as defined in Eq.~(\ref{newtrial}).}
\label{tab:parameters}
\end{center}
\end{table}

\subsection{Stark physics}\label{starkphysics}
Following our analogous treatment of the Stark effects of donors in Si, we modify the trial envelopes as defined in Eq.~(\ref{newtrial}) to account for the modifications that the ground wave function undergoes as a result of a uniform electric field that is turned on externally~\cite{PhysRevLett.94.186403}. This situation, which is common to all real-world settings and can provide a convenient way to control the qubit states, requires the inclusion of a term into the bulk Hamiltonian $H_{0}$ in Eq.~(\ref{eqbase}): $H=H_{0}-e \textbf{E}\cdot\textbf{r}$. We describe the corresponding modified $[111]$ envelope for illustration. The change in the envelope depends on the direction of the field relative to the principal axis of the effective mass tensor; if we define $\{x,y,z\}$ Cartesian axes along $\{[1\bar{2}1],[10\bar{1}],[111]\}$: 
\begin{flalign}\label{newenvelope}\nonumber
F_{111}= &N \left[\text{e}^{-\sqrt{\frac{x^{2}+y^{2}}{a^{2}_{s}}+\frac{z^{2}}{b^{2}_{s}}}} + \beta \hspace{1mm} \text{e}^{-\sqrt{\frac{x^{2}+y^{2}}{a^{2}_{l,{\hat {\bf E}}}}+\frac{z^{2}}{b^{2}_{l, {\hat {\bf E}}}}}}\right] \times \\  &\times (1+ q_{\hat{\bf E}} \hat{\bf E} \cdot {\bf r}) 
\end{flalign}
where $\hat {\bf E}$ is a unit vector in the direction of the applied field.
We now use an extra variational coefficient $q_{\hat {\bf E}}$, and allow the long range lengths $a_{l,{\hat {\bf E}}}$ and $b_{l,{\hat {\bf E}}}$ to change from their zero field values. With no further introduction of adjustable parts in our bulk pseudopotential, the expectation value of the energy is minimized over the complete singlet $A_{1}$ ground state. 

The angle of the applied field determines the degree to which the leftover bulk symmetry of the crystal is broken, and different angles lead to very different valley redistributions within the ground state. Let us consider the Hamiltonian matrix elements in the valley basis $\{\xi_{111},\xi_{\bar{1}11},\xi_{1\bar{1}1},\xi_{11\bar{1}}\}$: if $\textbf{E}\parallel \langle100\rangle$, the field orientation forms equal angles with all valleys, and so the Hamiltonian matrix takes the form $H_{\mu\nu}=\Lambda \delta_{\mu, \nu}+\Delta (1-\delta_{\mu, \nu})$, with $\delta_{\mu, \nu}$ the Kronecker delta function. The diagonal (intra-valley) entries $\Lambda$ and the off diagonal (inter-valley) couplings $\Delta$ are rigidly shifted from their zero-field values, but the overall symmetry of the resulting eigenstates is maintained: in fact, the ground state is still an equal superposition of all valleys, with eigenvalue $\Lambda+3\Delta$, while the excited states still form a triplet with energy $\Lambda-\Delta$.

At the opposite extreme, if $\textbf{E}\parallel [111]$, the field orientation makes a maximal distinction between the $[111]$ (aligned `$a$') valley and the other three (misaligned `m') valleys, which results in a Hamiltonian matrix of the form
\begin{equation} \label{matrix}
H=\left(
\begin{array}{cccc}
\Lambda_{a} & \Delta_{a} & \Delta_{a} & \Delta_{a} \\
\Delta_{a} & \Lambda_{m} & \Delta_{m} & \Delta_{m} \\
\Delta_{a} & \Delta_{m} & \Lambda_{m} & \Delta_{m}\\
\Delta_{a} & \Delta_{m} & \Delta_{m} &  \Lambda_{m} \\
\end{array}
\right),
\end{equation}
with $\Lambda_a=\langle \xi_{111}|H|\xi_{111}\rangle$, and, for example, $\Lambda_m=\langle \xi_{\bar{1}11}|H|\xi_{\bar{1}11}\rangle$, $\Delta_{a}=\langle \xi_{\bar{1}11}|H|\xi_{111}\rangle$, and $\Delta_{m}=\langle \xi_{\bar{1}11}|H|\xi_{\bar{1}11}\rangle$. The eigensystem now includes two singlets $G, E$ (ground, excited), with energies
\begin{flalign}
\label{diagonalization}
\epsilon_{G}^{E}=&\frac{1}{2}\biggl( 2\Delta_{m}+\Lambda_{a}+\Lambda_{m}\pm \\ & \pm |\sqrt{12 \Delta_{a}^2+(2\Delta_{m}-\Lambda_{a}+\Lambda_{m})^2}|\biggr)\nonumber,
\end{flalign}
and eigenvectors' coefficients
\begin{equation}
\label{alpha}\{\alpha_{\mu}\}^{E}_{G}=\frac{(\gamma_{G}^{E},1,1,1)}{\sqrt{3+ (\gamma_{G}^{E})^{2}}},
\end{equation}
where
\begin{flalign}
\nonumber
\gamma_{G}^{E}=&-\frac{1}{2 \Delta_{a} }\biggl[2\Delta_{m}-\Lambda_{a}+\Lambda_{m}\mp \\ & |\sqrt{12 \Delta_{a}^2+(2\Delta_{m}-\Lambda_a+\Lambda_m)^2}|\biggr];
\end{flalign}
and a doublet with intermediate energy $\Lambda_{m}-\Delta_{m}$, that is not Stark shifted.

The strong anisotropy of the Ge lowest conduction band ($m^\ast_\perp\approx20 m^\ast_\parallel$) implies that the squeezing of the state $\xi_i$ of each valley will be larger the more orthogonal $\hat{E}$ is to the valley axis. However, the overall Stark effect on the whole wave function results from the interplay of the squeezings of all valleys, each weighted with the coefficients in Eq.~(\ref{alpha}). We discuss more quantitatively these features in the next section, when we show theoretical predictions, backed up by some experimental measurements, for Stark shifts of Ge:P, Ge:As and Ge:Bi in different geometries of applied electric and magnetic fields. 

In the regime of low electric fields relevant to the recent experiments ($|{\bf E}|\sim$kV/m) of Ref.~\onlinecite{sigillito16}, the intervalley couplings $\Delta$ are shifted much less than the diagonal valley energies $\Lambda$, as they involve higher Fourier components of the linear electric-field term in the Hamiltonian. While all terms will be retained in the numerical calculations that follow, it is instructive to set $\Delta_m\approx\Delta_a\approx\Delta_{0}$ ($\Delta_{0}$ being the zero-field intervalley coupling), and acknowledge that owing to the large anisotropy  described above $|\Lambda_a-\Lambda_0|\ll |\Lambda_m-\Lambda_0|$ (where $\Lambda_0$ the zero field intravalley coupling). If we now expand up to second order in the magnitude of the applied electric field, we find:
\begin{flalign}\label{eshift}
\epsilon_{G}(\textbf{E}\neq0)-\epsilon_{G}(\textbf{E}=0)&\approx \frac{3}{4}(\Lambda_m-\Lambda_a),\\
\label{gamma}
\gamma_{G}&\approx 1-\frac{1}{4}\frac{(\Lambda_m-\Lambda_a)}{\Delta_0},
\end{flalign} 
The quadratic dependence of these quantities on $|\textbf{E}|$ is thus made apparent, as linear terms in $\Lambda$ are forbidden by the parity symmetry within each valley~\cite{PhysRevB.90.195204}. Furthermore, the reorganization of the valleys that is represented by the coefficient $\gamma_{G}$ [see Eq.~(\ref{alpha})] can be now understood as the result of the interplay of the anisotropy of the Ge conduction band, that dictates how the longitudinal and transverse valleys envelopes respond to the applied field within each valley, and the inter-valley coupling $\Delta_0$ (that sets the singlet-triplet energy difference). These features will be shown to fundamentally determine the large spin-orbit Stark shifts characteristic of Ge donors, as further detailed in Section~\ref{spinorbit}.

\section{Single-qubit manipulation}\label{singlequbit}
The adjustment of the donor wave function to local electrostatic distortions, as described so far, has important consequences on the donor spin resonance spectrum, as it affects both the contact hyperfine coupling $A\textbf{I}\cdot\textbf{S}$ between the nuclear spin $I$ and the electron spin $S$, and the $\textbf{g}$-tensor that modulates the Zeeman interaction of the electron spin with an external magnetic field. The chance to tune the Hamiltonian of donor spins in a controllable way provides natural methods of qubit selective manipulation~\cite{kane98}, that have been largely investigated with Si donors~\cite{wave functioncontrol,PhysRevLett.97.176404,PhysRevLett.113.157601,PhysRevLett.114.217601,Lauchte1500022}. With a view to exploring similar applications with Ge donors, it is important to quantify the possible range of these tunings in a realistic regime of applied electric and magnetic fields. 

\subsection{Hyperfine Stark shifts}

For electric fields well below ionization, the relative hyperfine Stark shifts can be parametrized by a single quadratic coefficient $\eta_{a}$, the `hyperfine susceptibility':
\begin{equation}\label{etaaprec}
\frac{\Delta A}{A_{0}}=\frac{|\Psi(\textbf{E}\neq 0,\textbf{r}=\textbf{0})|^2}{|\Psi(\textbf{E}= 0,\textbf{r}=\textbf{0})|^2}-1\equiv \eta_{a} E^2.
\end{equation}
Direct substitution of the Stark-shifted donor wave function discussed in the previous section, for different orientations of the applied $\textbf{E}$, in Eq.~(\ref{etaaprec}) leads to  
\begin{flalign}\label{etaa}\nonumber
\frac{\Delta A}{A_0}\langle111\rangle=&\frac{1}{4F_0^2(\textbf{0})}\frac{1}{3+\gamma_{G}^2}(3 F_{\bar{1}11}(\textbf{0})+\gamma_{G} F_{111}(\textbf{0}))^{2}-1, \\
\frac{\Delta A}{A_0}\langle100\rangle=&\frac{1}{F_0^2(\textbf{0})}F_{111}^2(\textbf{0})-1, \\
\frac{\Delta A}{A_0}\langle110\rangle=&\frac{1}{4F_0^2(\textbf{0})}\frac{1}{2+2\gamma_{G}^2}(2 F_{\bar{1}11}(\textbf{0})+2\gamma_{G} F_{111}(\textbf{0}))^{2}-1,\nonumber
\end{flalign}
where $F_{0}(\textbf{0})$ is the amplitude of any of the envelopes defined in Eq.~(\ref{newtrial}), evaluated at the impurity site. 

Our theoretical predictions of the $\eta_{a}$ coefficients are reported in Table~\ref{tab:starkhyperfine} for all donors and the three different field orientations considered here. Eqs.~(\ref{etaa}) clarify the two different effects of an external field: the envelopes of each valley are renormalized, and the valleys are repopulated as described by the shift of the coefficient $\gamma_{G}$ from its zero-field value $\gamma_{G}=1$. Both effects, as described by Eqs.~(\ref{eshift}) and~(\ref{gamma}), give a direct measure of how much the electric field distinguishes between two different valleys. The first effect makes the largest contribution to the total susceptibility $\eta_a$, but the second is more responsible for the dependence on the field orientation, which is more pronounced for weaker valley-orbit effects [see Eq.~(\ref{gamma})]. As a consequence, the hyperfine susceptibilities only change mildly with the orientation of the field, but for Ge:P $\Delta_0$ is small enough that $\eta_a$ can vary by $\sim 10\%$ between different field directions.  

Direct comparison of Table~\ref{tab:starkhyperfine} with experiments performed with Ge:P and Ge:As donors in Ref.~\onlinecite{sigillito16} shows very close agreement in both cases.

The hyperfine susceptibilities we have calculated are about two orders of magnitude larger than those of Si donors. As the bulk hyperfine constants $A_0$ of Ge:P and Ge:As are about half than the respective Si:P and Si:As donors, and for Ge:Bi it is about five times smaller than in Si:Bi, the absolute hyperfine shifts are 50 or 20 times larger if the same electric field is applied. The interesting question to pose for considering whether hyperfine-shifted Ge donors could be more tunable qubits than those in Si requires comparing the maximum shifts $\Delta A$ that could be attained. In fact, as Ge donor wave functions are more spread out in real space, it is expected that they will also more easily completely lose the contact hyperfine coupling, which happens when the field is strong enough that $p$ orbital states, with negligible density at the impurity site, are strongly admixed in the Stark shifted ground state. To determine when this effect comes into play, we can calculate the
the Stark shifted $2p$ state can be calculated within the same theory developed here, once a suitable $2p$-like trial envelope is introduced:~\cite{PhysRevB.90.195204}
\begin{equation}
\Psi^{2p}(\textbf{r})=N_{p}\hspace{1mm} z \hspace{1mm}e^{-\sqrt{\frac{x^2+y^2}{a_p^2}+\frac{z^2}{b^2_p}}}(1+q^p_z z).
\end{equation}
We find that the binding energy of the $2p$ manifold for all donors approaches 10 meV, which is close to the $1s_{A1}$ ground states' energy, when the applied field is of the order of $0.2$~MV/m. Since the $1s_{A1}$ energies do not move significantly from their bulk value in this regime, at this point the $2p$ orbital manifold will anticross the $1s_{A1}$ state for all donors, and the adiabatic ground state will contain a significant amount of $p$-character. As a result, for $E\approx 0.2$MV/m, $\Delta A$ will no longer be described by the quadratic susceptibility $\eta_a$, but will vary much more strongly with $E$, giving an unappealing unstable spin energy splitting. Thus, if the qubit states are split by the hyperfine interaction, the maximum stable $\Delta A$ that can be achieved with bulk donors are of the order of $0.27\times(0.2)^2\times58.5$~MHz$~=0.63$~MHz (Ge:P), $0.12\times(0.2)^2\times 99.3$~MHz~$=0.48$~MHz (Ge:As), $0.17\times(0.2)^2\times292$~MHz~$=$~1.99MHz (Ge:Bi), corresponding to maximum ESR frequency shifts of 0.3 MHz, 1 MHz, and 9.5 MHz: these values are curiously very close to those calculated for Si donors in Ref.~\onlinecite{PhysRevB.90.195204}. We conclude that gate times for selective manipulations of spin qubits relying on resonant excitation of locally Stark shifted donors in the bulk are not improved from Si to Ge.

\begin{table} 
\begin{tabular}{l@{\hspace{10pt}} *{3}{c}}
\hline
\textbf{Donor} & $\eta_{a} (\mu m^{2}/V^{2})$ (th)  & $\eta_{a} (\mu m^{2}/V^{2})$ (exp) \\
$\hat{\textbf{E}}$ & $\langle111\rangle \hspace{3mm} \langle100\rangle \hspace{3mm} \langle110\rangle$ & \\
\hline
P & -0.27 \hspace{3mm} -0.24 \hspace{3mm} -0.21& -0.22 \\
\hline
As & -0.12\hspace{3mm} -0.1 \hspace{3mm} -0.1 & -0.12\\
\hline
Bi & -0.17\hspace{3mm} -0.17 \hspace{3mm} -0.17 & /\\
\hline
\end{tabular}
\caption{Comparison of the theoretical quadratic Stark shift coefficients $\eta_{a}$(th) of the hyperfine couplings of three group V donors in Ge, as calculated from Eq.~(\ref{etaa}), and the corresponding experimental values $\eta_{a}$(exp) measured in Ref.~\onlinecite{sigillito16}. }
\label{tab:starkhyperfine}
\end{table} 
 
\subsection{Spin-orbit Stark shifts}\label{spinorbit}
If an applied electric field is also combined with an external magnetic field, the donor response includes a so-called spin-orbit Stark effect. Following Ref.~\onlinecite{feher}, let us define the $\textbf{g}$ tensor as the response of the donor electron spin $\textbf{S}$ to some external magnetic field $\textbf{B}$:
\begin{equation}\label{gfactor}
H_{mag}=\textbf{S}\cdot \textbf{g} \cdot \textbf{B}.
\end{equation}
The spin-orbit interaction of conduction electrons with the host Ge nuclei is strong enough that the bulk $g$-factor $g_0=1.57$ differs strongly from the free electron value $g\sim2$. Due to the large mass anisotropy of the Ge conduction band, each valley axis $\hat{\mu}$ displays two distinct $g$-tensor components depending on whether $\textbf{B}$ is parallel ($g_{\parallel}$) or perpendicular ($g_{\perp}$) to it. For a general $\textbf{B}$-orientation angle $\theta$ with respect to any valley-axis $\hat{\mu}$, the measured single-valley $g_{\mu}$-factor is~\cite{feher}
\begin{equation}\label{scalarg}
g_{\mu}^{2}\equiv \frac{|\textbf{g}_{\mu}\cdot\textbf{B}|^2}{|\textbf{B}^2|}=\sum_{i,j} (g_{\mu}^{ij}B^{j})^2/|\textbf{B}^2|=g_{\parallel}^{2} \cos^{2}{\theta} + g_{\perp}^{2}\sin^{2}{\theta},
\end{equation}
where $i$ and $j$ run over the three Cartesian components of the tensors. The overall $\textbf{g}_0$-tensor in bulk donors is an equal superposition of all the four valley $\textbf{g}_\mu$-tensors. Experimental measurements in Ref.~\onlinecite{wilson} show that $g_{\perp}=1.92 \pm 0.02$, $g_{\parallel}=0.87 \pm 0.04$, with relative values across different implanted donor species only varying by a few percent. 

In the presence of applied gate voltages or strain perturbations that break the tetrahedral symmetry of the doped crystal, the overall bulk $\textbf{g}=\sum_{\mu}\alpha_{\mu}\textbf{g}_{\mu}$ can either be tuned by single-valley modifications of each $\textbf{g}_\mu$ or by the valley repopulation that reorganizes the $\alpha_\mu$ coefficients~\cite{feher}.
Due to the peculiar features of Ge donor states, we will show that these variations can be very large, thus providing a very effective way to electrically manipulate donor qubit states -- much more appealing than the hyperfine detuning examined in the previous subsection.

Single-valley $g_{\mu}$-values can only be shifted by an applied electric field that is not along any $\langle 111 \rangle$ crystallographic direction -- this was pointed out by Roth~\cite{wilson} and experimentally verified in the same work. In this case, the shift is due to the field enhancing the spin-orbit coupling of other bands to the lowest conduction band. On the other hand, valley repopulation occurs when $\textbf{E}$ is not parallel to any $\langle 100 \rangle$ direction (i.e. when $\textbf{E}$ does not make the same angle with all valleys): this effect is due entirely to the Stark physics investigated in Sec.~\ref{starkphysics}. The relative importance of these two kinds of orbital coupling can be preliminarily assessed by comparing the energy differences between the respective unperturbed states to be coupled: the inter-band energy difference involved in the single-valley mechanism is two orders of magnitude larger than the $E_{T2}-E_{A1}$ splitting~\cite{roth}, which is relevant for valley repopulation. It is thus predicted that the $g$-factor shifts will be much larger when the valley repopulation effect plays a role, and this has been indeed verified by \emph{ab initio} calculations in Ge:P~\cite{PhysRevB.80.155301}. Thus, we limit ourselves to the calculation of the more important spin-orbit Stark shifts induced by valley repopulation, 
especially since approach relies on a single band approximation which is unable to account for the single valley $g$-factor shifts. 
As a consequence, our theoretical predictions aim at describing the full $g$-factor shift if $\textbf{E}\parallel\langle111\rangle$, giving a very tight lower bound if $\textbf{E}\parallel\langle110\rangle$, while the smaller shifts in the case $\textbf{E}\parallel\langle100\rangle$ cannot be described. 


From the discussion in Sec.~\ref{starkphysics}, we know that the valley repopulation is most effective when $\textbf{E}\parallel\langle111\rangle$, as the valleys are maximally distinguished in this case. In the Cartesian frame $\{[100], [010], [001]\}$, substitution of the valley coefficients $\alpha_{\mu}$ from Eq.~(\ref{alpha}) gives $\textbf{g}_{i,j}=\sum_{\mu}\alpha_{\mu}\textbf{g}^{\mu}_{i,j}=g_{0}\delta_{i,j}+\Delta_{g}(1-\delta_{i,j})$, where $g_{0}=g_{\parallel}/3+2g_{\perp}/3$ is the average bulk $g$-factor, and
\begin{equation}\label{deltag}
\Delta_{g}\equiv \frac{g_{\parallel}-g_{\perp}}{3}(\gamma_{G}^2-1)/(3+\gamma_{G}^2).
\end{equation}

The analytical expression for $\gamma_{G}$ in Eq.~(\ref{gamma}) immediately shows that the spin-orbit Stark shift of bulk donors depends quadratically on $\textbf{E}$. Different directions of the applied magnetic field $\textbf{B}_{0}$ will excite different combinations of the $\textbf{g}$ tensor components [in the sense clarified by Eq.~(\ref{scalarg})]: for example, if $\textbf{B}_{0}\parallel\langle111\rangle$, then simple matrix multiplication gives for the scalar $g$-factor $g=g_{0}+2 \Delta_{g}$, while if $\textbf{B}_{0}\perp\langle111\rangle$ then $g=g_{0}- \Delta_{g}$. 

\begin{figure}[t!]

\centering
\includegraphics[width=.45\textwidth]{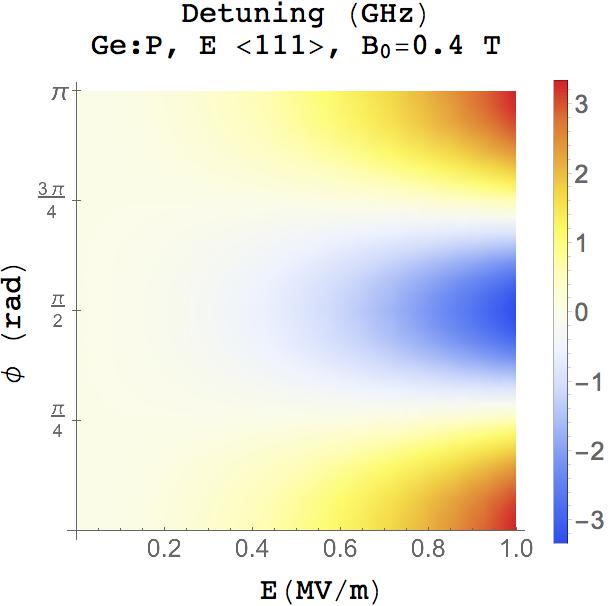}
\caption{Density plot of the Stark-detuned Zeeman coupling $[g(E)-g_0] \mu_{\rm B} B_{0} $ induced by a nonzero electric field \textbf{E}, aligned with a $\langle111\rangle$ crystallographic direction, with an angle $\phi$ between $\textbf{E}$ and the magnetic field $\textbf{B}_0$. For reasonable values of applied voltage, the detuning can be changed over the range $\{- 3$ \text{GHz}, 3 $\text{GHz}\}$, allowing for nanosecond selective manipulation of locally detuned spins. The symmetry for reflections $\phi\rightarrow(\pi/2-\phi)$ is a consequence of the cylindrical symmetry of this field configuration: the electric field is parallel with a valley axis. As explained in the text, in this geometry the largest range of detunings can be attained. $B_0=0.4$~T corresponds to a donor spin coherence $T_2\approx1$~ms, as measured in Ref.~\onlinecite{PhysRevLett.115.247601}.}
\label{fig:gfac}
\end{figure}

Remarkably, the detuning $(g-g_0)\mu_{\rm B}B_0$ (where $\mu_{\rm B}$ is the Bohr magneton) can be very large: Fig.~\ref{fig:gfac} shows a density plot of the detuning of a Ge:P bulk spin with $\textbf{E}\parallel\langle111\rangle$ and $B_0=0.4$~T (corresponding to $T_2\approx1$~ms as measured in Ref.~\onlinecite{PhysRevLett.115.247601}) as a function of the electric field magnitude and the angle between $\textbf{E}$ and $\textbf{B}_0$. Strikingly, we show that qubit detunings above GHz could be attained within realistic experimental settings, thus allowing for nanosecond selective resonant manipulation -- a two-orders-of-magnitude improvement to the maximum speed achievable with detuned Si donor spins.

After defining
\begin{equation}\label{etagdef}
\frac{g(E)}{g(0)}-1=\eta_{g}|\textbf{E}|^{2},
\end{equation}
in Table~\ref{tab:starkgfactor} we extend our predictions to Ge:As and Ge:Bi, and consider the further situation $\textbf{E}\parallel \langle110\rangle$. This latter case is characterized by smaller valley repopulation, as explained in Sec.~\ref{starkphysics}, and accordingly shows smaller spin-orbit Stark shifts. 

Our analytical framework highlights clearly the sources of this giant tunability: combining Eqs.~(\ref{deltag}) and~(\ref{gamma}), we can rewrite
\begin{equation}\label{deltagex}
\Delta_{g}= \frac{g_{\parallel}-g_{\perp}}{3}\frac{\epsilon_{G}-\epsilon_{G}^{0}}{6|\Delta_{0}|},
\end{equation}
hence very large $\Delta_g$ are a consequence of the large mass anisotropy in the Ge conduction band [which leads to large $\epsilon_G-\epsilon_G^0$, as seen from Eq.~(\ref{eshift})] and the relatively small valley-orbit couplings $\Delta_0$. In Si donors, the weaker spin-orbit interaction and the smaller anisotropy make the term $g_{\parallel}-g_{\perp}$ three orders of magnitude smaller, and a reduction of another order of magnitude comes from the donors being less shallow (as a consequence of larger valley-orbit interaction): this is indeed compatible with measurements in Ref.~\onlinecite{PhysRevLett.114.217601}.

Eq.~(\ref{deltagex}) also clearly shows that $\Delta_g>0$, since we find that the donor is more bound with increasing magnitude of the applied electric field, but $g_\parallel<g_\perp$ -- this is indeed confirmed by recent measurements of Ge:As in Ref.~\onlinecite{sigillito16}. More remarkably, the agreement between this work and our theory is very good, as shown by Table~\ref{tab:starkgfactor}: this validates our theory and builds our confidence in all the results and the predictions presented in this paper.

\begin{table} 
\begin{tabular}{l@{\hspace{10pt}} *{4}{c}}
\hline
\textbf{Donor} & \textbf{B}$_{0}$ orientation & $\eta^{\text{th}}_g$ ($\mu$m$^{2}/$V$^2$)   & $\eta^{\text{exp}}_g$  ($\mu$m$^{2}/$V$^2$)   \\
$\hat{\textbf{E}}$ & &$\langle111\rangle$  \hspace{3mm}  $\langle110\rangle$ & $\langle111\rangle$  \hspace{3mm} $\langle110\rangle$ \\
\hline
P & \textbf{B}$\parallel$\textbf{E} & 0.19 \hspace{3mm} 0.1 & /\\
\hline
P & \textbf{B}$\perp$\textbf{E} & -0.095 \hspace{3mm} -0.1 & /\\
\hline
As & \textbf{B}$\parallel$\textbf{E} & 0.04 \hspace{3mm} 0.017 &0.04 \hspace{3mm}0.017 \\
\hline
As & \textbf{B}$\perp$\textbf{E} & -0.02 \hspace{3mm} -0.017 & -0.03\hspace{3mm} -0.012\\
\hline
Bi & \textbf{B}$\parallel$\textbf{E} & 0.03 \hspace{3mm}  0.012 &/\\
\hline
Bi & \textbf{B}$\perp$\textbf{E} & -0.015\hspace{3mm} -0.012 &/\\
\hline
\end{tabular}
\caption{Comparison of the theoretical quadratic Stark shift coefficients $\eta_{g}$(th) of the g-factors of three group V donors in Ge, as calculated from Eq.~(\ref{deltag}), and the corresponding experimental values $\eta_{g}$(exp) measured in Ref.~\onlinecite{sigillito16}. }
\label{tab:starkgfactor}
\end{table} 

\section{Two-qubit coupling}

The exchange interaction $J$ between neighboring electron spins with overlapping orbital densities provides the most natural spin qubit coupling in semiconductor architectures, with detailed manipulation schemes that have been proposed\cite{kane98} and demonstrated e.g. with quantum dots\cite{Veldhorst2015}. The feasibility of two-qubit gates is another key figure of merit of candidate qubits, thus it is critical to understand how large inter-spin couplings can be, and how robust they are in the face of realistic fabrication defects. Direct comparison of Table~\ref{tab:parameters} with our previous work\cite{PhysRevB.90.195204} shows that donor wave functions are indeed much larger in Ge than in Si, hence it is expected that the coupling strength between two donor spins separated by the same distance will be much larger it is in Ge, since it increases exponentially with effective Bohr radius. However, Coulomb interactions of electron states in multi-valley semiconductors are known to depend strongly, with sub-nanometer resolution, on the relative positions of the two coupled spins; since state-of-the-art experimental control on donor placement is limited to precision of a few nm\cite{lansbergen2008,Pierre2009,Fuechsle2012}, all the pairs within a qubit cluster implanted in a realistic device will experience very different couplings. In this section, we therefore wish to quantify the donor distance constraints needed for significant coupling between Ge:P spins, and to compare realistic distributions of such couplings across the two semiconductors, Si and Ge.  

We start with some qualitative considerations that are common to $J$ couplings in multi-valley semiconductors. The Heitler-London approximation\cite{hl} can give reliable estimates of the exchange splitting $J=E_{T}-E_{S}$ between the singlet $\ket{S}=\frac{1}{\sqrt{2}}\ket{\uparrow\downarrow-\downarrow\uparrow}$ and triplet $\ket{T_{0,+,-}}=\frac{1}{\sqrt{2}}\ket{\uparrow\downarrow+\downarrow\uparrow},\ket{\uparrow\uparrow},\ket{\downarrow\downarrow}$ spin states of two neutral donors interacting via the Heisenberg antiferromagnetic interaction $J \textbf{S}_{1}\cdot\textbf{S}_{2}$, with $J>0$ at zero magnetic field. If the overlap between two single-electron wave functions centered at different sites $\textbf{R}_{a}, \textbf{R}_{b}$, namely ${\cal S}=\langle \Psi_{a}(\textbf{r}-\textbf{R}_{a})|\Psi_{b}(\textbf{r}-\textbf{R}_{b})\rangle$, is small enough, the Heitler-London prescription leads to an estimate of $J$ based on a combination of single and two-particle integrals over the isolated wave functions of the non-interacting system\cite{oscilla,oscilla2}. The reliability of such calculations decreases if the overlap is too large, i.e. if the donors are too close. It also fails if the donors are too far apart -- $|\textbf{d}=\textbf{R}_{a}-\textbf{R}_{b}|$ larger than about 50 effective Bohr radii -- where it is found that $J$ can turn negative, thus contradicting the Lieb-Mattis theorem\cite{mattis}. However, we are interested in the intermediate regime, is a compromise between significant qubit coupling and the technical difficulties of fabrication very close donors. Thus, the Heitler-London approach has commonly been taken in theoretical predictions of donor couplings in the last fifteen years\cite{oscilla,oscilla2,wandh,PhysRevB.90.195204}. The successive advances in understanding and the improvements of calculations of donor couplings have only come from improvements to the theory of the single donor wave function. We remark that all of the integrals contributing to $J$ depend crucially on how the wave function behaves in the spatial region between the two donors, i.e. far enough from the central cell, that our effective mass treatment, calibrated as in Sec.~\ref{mvemt} and verified to agree with experiments in Sec.~\ref{singlequbit}, should provide reliable predictions in the regime of interest.

The analytical Heitler-London expression for $J$ can be combined into four different terms sharing the same valley structure (see Appendix A), and thus showing similar dependences on the donor separation. For the sake of illustration, let us consider the simplest, namely the overlap integral squared ${\cal S}^2=\langle\Psi(\textbf{r})|\Psi(\textbf{r}-\textbf{d})\rangle$:
\begin{equation}\label{overlap}
{\cal S}^2=\left(\sum_{\mu,\nu}\alpha_{\mu}\alpha_{\nu}\langle \xi_{\mu}|\xi_{\nu}\rangle\right)^2\approx \left(\frac{1}{\symbol{35} \text{valleys}}\sum_{\mu}e^{-i \textbf{k}_{\mu}\cdot\textbf{d}}{\cal S}_\mu\right)^2
\end{equation} 
where ${\cal S}_\mu\approx\int d\textbf{r} F_{\mu}(\textbf{r}-\textbf{R}_a) F_{\mu}(\textbf{r}-\textbf{R}_b)$ is the overlap between envelopes with the same spatial phase, i.e. those pertaining to the same valley. The terms in Eq.~(\ref{overlap}) with mixed valleys $\mu\neq\nu$ are suppressed by the large spatial oscillations that come from the phase differences $(\textbf{k}_{0\mu}-\textbf{k}_{0\nu})\cdot\textbf{r}$. In the expression of ${\cal S}_\mu$ only the in-phase plane waves of the Bloch functions in Eq.~(\ref{waveexpansion}) have been retained, which is again an excellent approximation due to the envelopes not changing significantly over the lengthscale $1/k_0$. Since time-reversal symmetry in the Ge crystal guarantees that $F_{\mu}=F_{-\mu}$, we can further manipulate Eq.~(\ref{overlap}) into
\begin{equation}\label{overlapsimple}
{\cal S}^2\approx\frac{1}{(\symbol{35} \text{valleys})^2}\sum_{\mu,\nu} \cos[(\textbf{k}_{0\mu}-\textbf{k}_{0\nu})\cdot\textbf{d}]{\cal S}_\mu {\cal S}_\nu,
\end{equation}  
which highlights the importance of the structure of the conduction band for coupling two spins embedded in multi-valley semiconductors: just like ${\cal S}^2,$ $J$ will depend on the donor separation $\textbf{d}$ not only through the smoothly decaying envelope overlaps ${\cal S}_\mu$, but also through the highly oscillatory cosines $\cos[(\textbf{k}_{0\mu}-\textbf{k}_{0\nu})\cdot\textbf{d}]$ whose period is of the order of the crystal lattice constant. Practical implementation of two-qubit operations with uniform control will have to deal with the vast range of coupling magnitudes intrinsic to an ensemble of qubits with different donor separations $\textbf{d}$.  

\begin{figure}[t!]
\setlength{\belowcaptionskip}{-5pt}

\centering
\begin{tabular}{l}

\includegraphics[width=.45\textwidth]{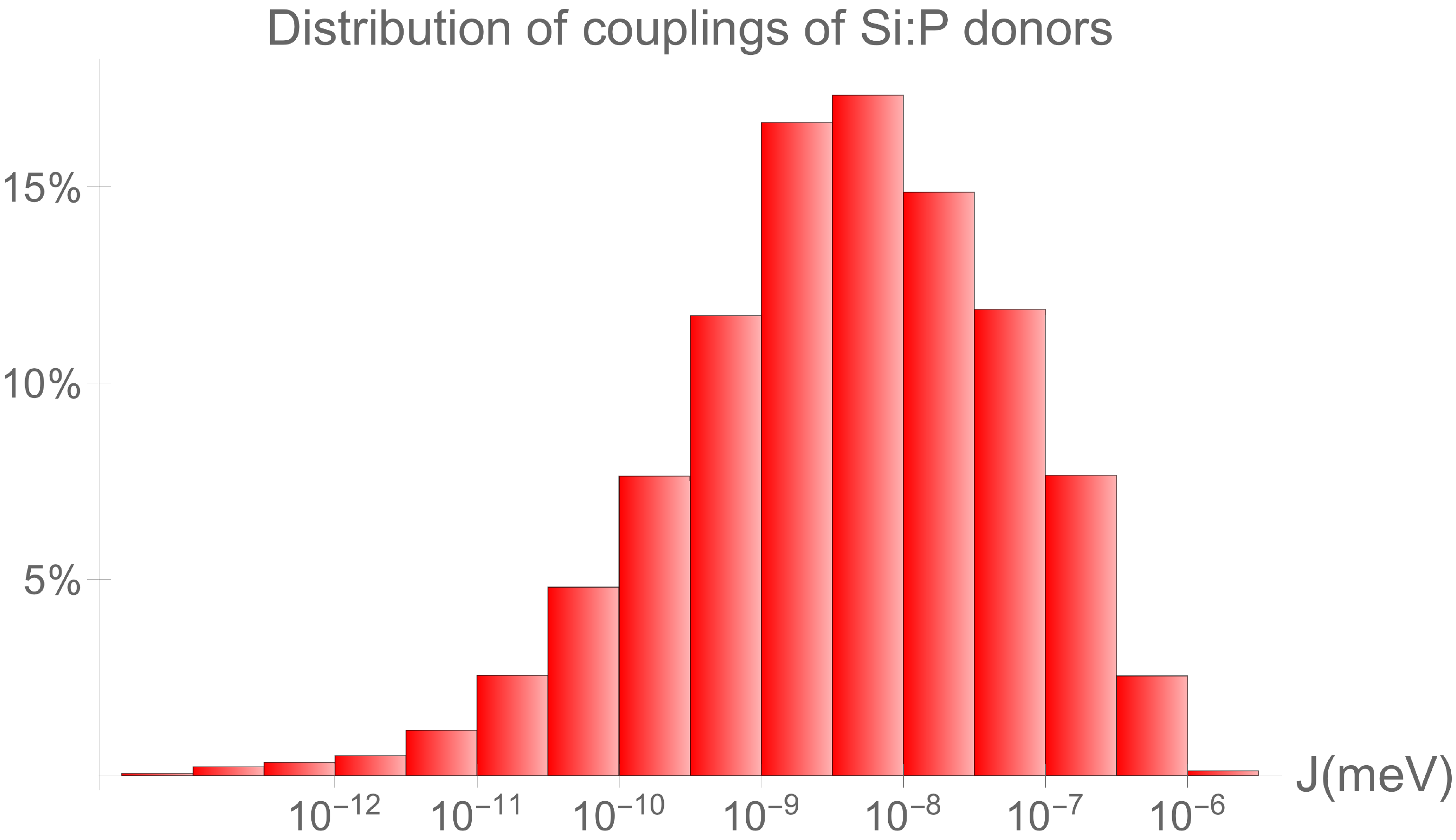}\llap{\makebox[7.5cm][l]{\raisebox{2.4cm}{\includegraphics[width=.17\textwidth]{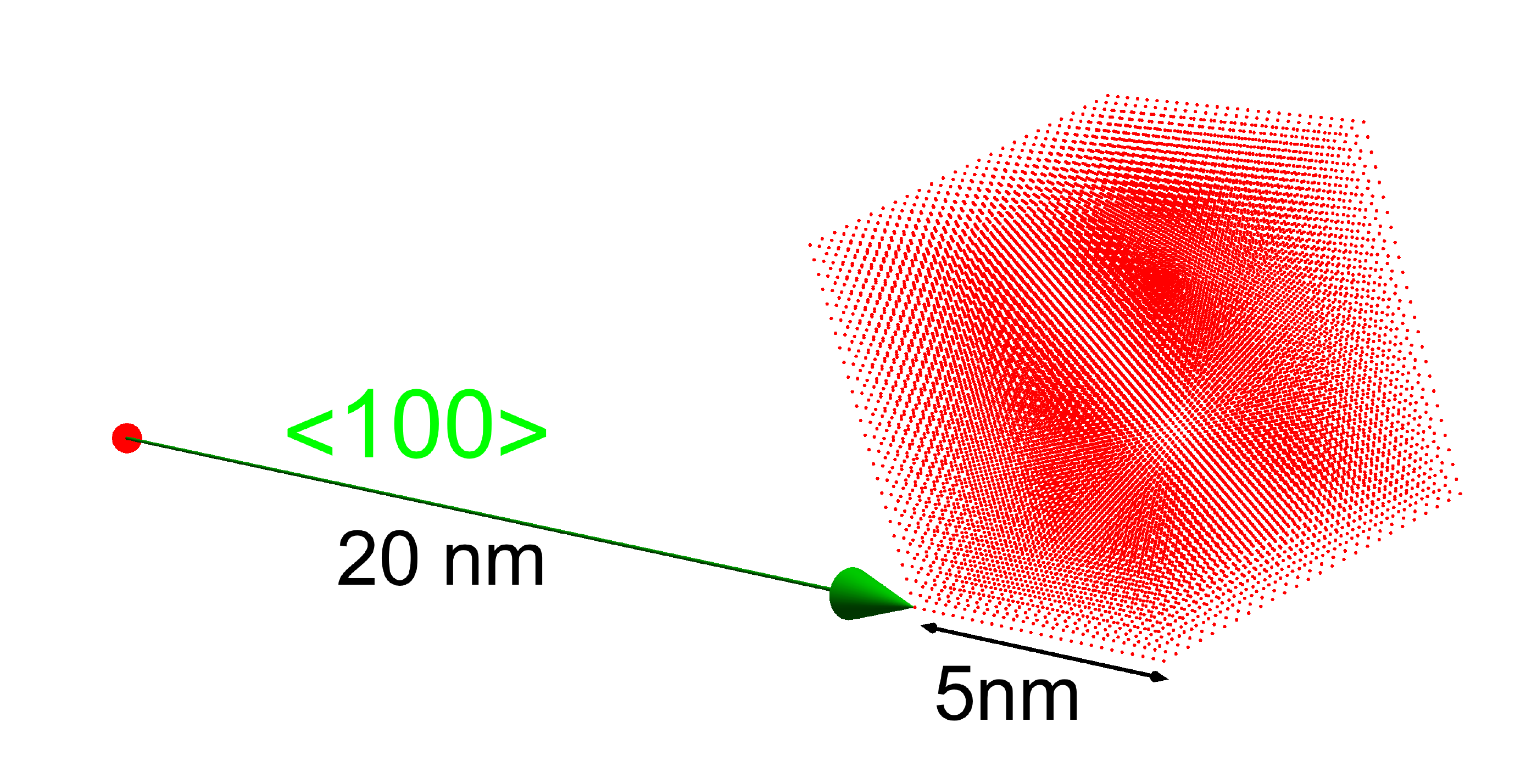}}}} \\
\\
\includegraphics[width=.45\textwidth]{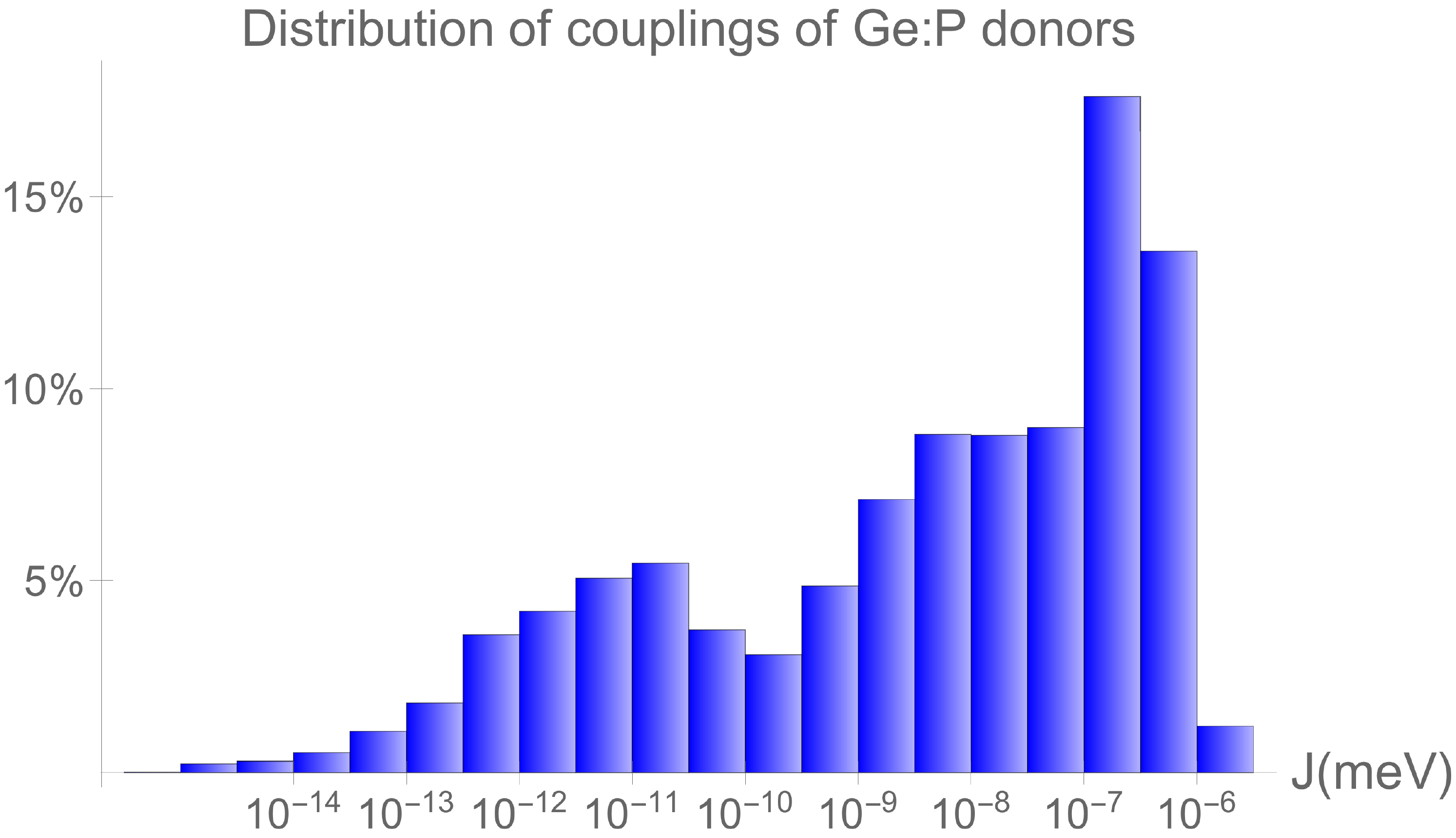}\llap{\makebox[7.5cm][l]{\raisebox{2.4cm}{\includegraphics[width=.17\textwidth]{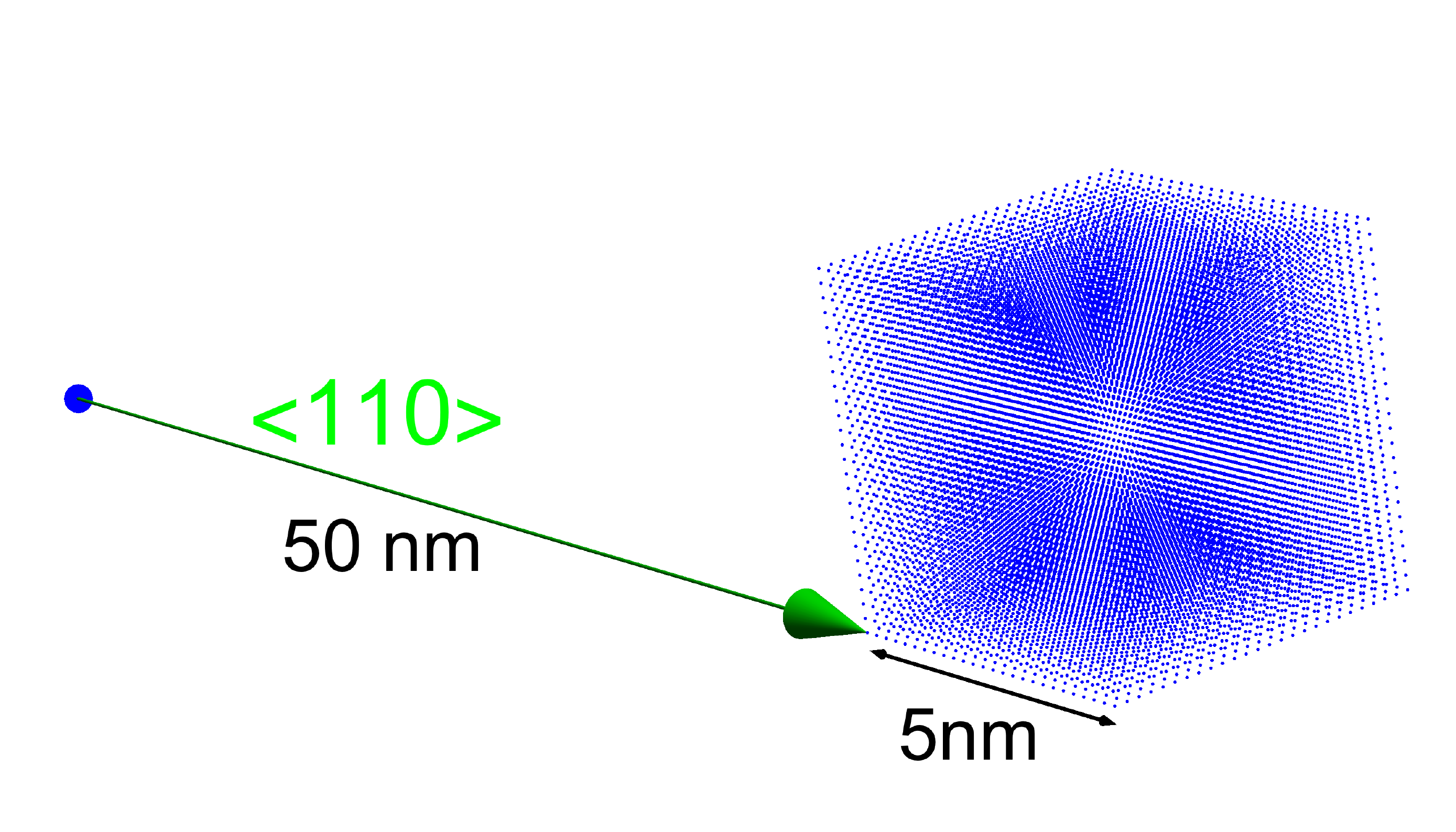}}}} \\
\end{tabular}
\caption{Distribution of $J$ couplings (on a log-scale) of a statistical ensemble of Si:P (top) and Ge:P (bottom) donor pairs, with pair separations $\textbf{d}_0+\textbf{d}_r$, where the nominal separations are $\textbf{d}^{\text{Si}}_0=[20 $ $\text{nm}, 0, 0]$, $\textbf{d}^{\text{\rm Ge}}_0=[50/\sqrt{2}$ $ \text{nm}, 50/\sqrt{2}$ $ \text{nm}, 0]$ for Ge and Si respectively. These combine with a random separation $\textbf{d}_r$ that is uniformly distributed over cubes with edge length 5 nm. These cubes are illustrated in the insets and show all the possible positions of donors partners that couple to a fixed donor at the origin of the $\textbf{d}_0$ vector. The Ge:P distribution spreads across more orders of magnitude overall, but it is more skewed towards larger couplings than that for Si:P. 
}
\label{fig:histos}
\end{figure}

The first question that we address regards the best geometry that shrinks as much as possible this spread of $J$ couplings: we set to model a statistical ensemble of donor pairs whose separation vectors $\textbf{d}$ are combinations of the same nominal distance $\textbf{d}_0$ and a random vector $\textbf{d}_r$ that is uniformly distributed in a cube with edge length of 5 nm. Thus we assume that, within each coupled pair of the ensemble, one donor is fixed at one vertex of the separation vector $\textbf{d}$, while the other can take any of the equally likely positions within the cube illustrated in the insets of Fig.~\ref{fig:histos}. This represents a realistic scenario compatible with the imprecision in donor positioning of modern ion implantation methods\cite{Schenkel2009} that are used to fabricate semiconductor nanodevices for quantum computing.

It is known\cite{oscilla2} that, as explained in Ref.~\onlinecite{PhysRevB.89.235306}, the best option for $\textbf{d}_0$ in Si is to lie along a $\langle100\rangle$ crystallographic direction. Due to the different band structure in Ge, we find that a different choice of $\textbf{d}_0$ is optimal in this case: this requires considering all the possible sets of $\{(\textbf{k}_{0\mu}-\textbf{k}_{0\nu})\cdot\textbf{d}_0\}_{\mu,\nu=1,2,3,4}$ when $\textbf{d}_0$ varies across the different crystallographic directions, plus taking into account the role of the anisotropy of the lowest conduction band. In fact, if the envelopes $F_{\mu}$ were spherically symmetric [so that ${\cal S}_{\mu}(\textbf{r})={\cal S}_{\nu}(\textbf{r})$ for all $\mu, \nu$], then the optimal $\textbf{d}_0$ minimizing the spread of ${\cal S}^2$ values in Eq.~(\ref{overlapsimple}) for the random misplacements $\textbf{d}_r$ would be the one that minimizes the sum $\sum^4_{\mu,\nu=1} \cos[(\textbf{k}_{0\mu}-\textbf{k}_{0\nu})\cdot\textbf{d}_0]$. Thus, $\textbf{d}_0$ should be as close to orthogonal as possible with all of the pairwise differences of valley vectors $\textbf{k}_{0\mu}$, i.e. it should make equal angles with all valleys, i.e. $\textbf{d}_0\parallel \langle100\rangle$. However, the very large anisotropy of the Ge conduction band implies that it is better to have as many terms as possible with $(\textbf{k}_{0\mu}-\textbf{k}_{0\nu})\perp\textbf{d}_0$, rather than distributing the orthogonality equally among all terms~\cite{PhysRevB.89.235306}: this is attained with $\textbf{d}_0\parallel \langle110\rangle$. 

\begin{figure}[t!]

\centering
\begin{tabular}{ll}
\includegraphics[width=.24\textwidth]{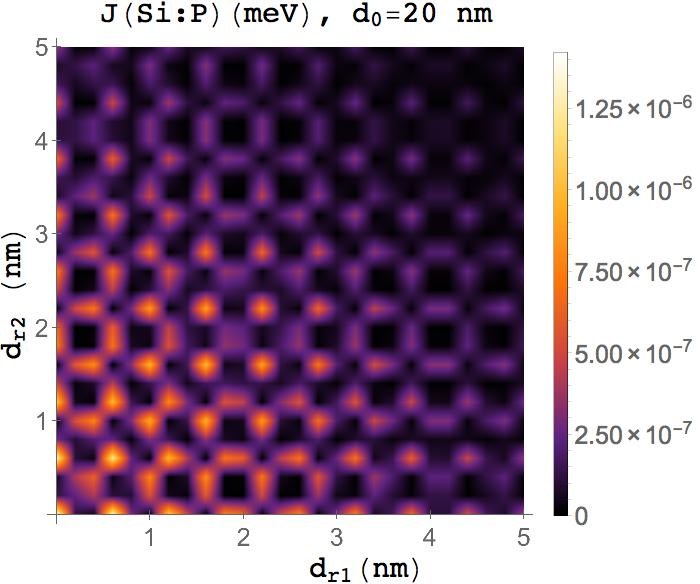} & \includegraphics[width=.24\textwidth]{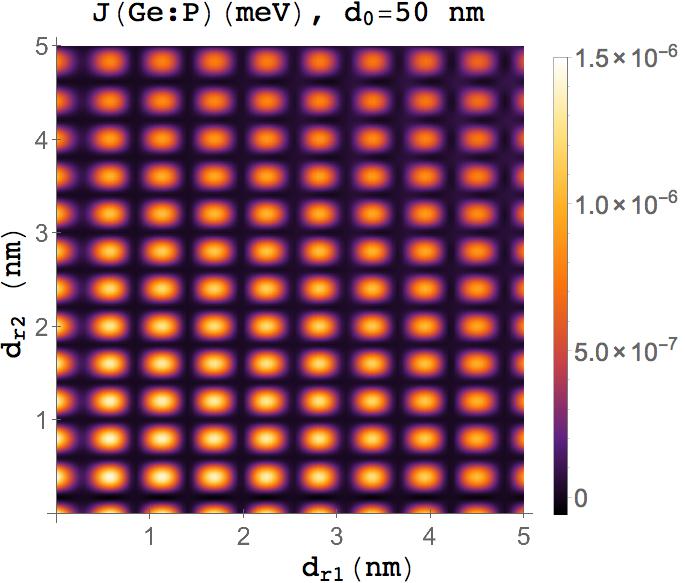}\\
\includegraphics[width=.24\textwidth]{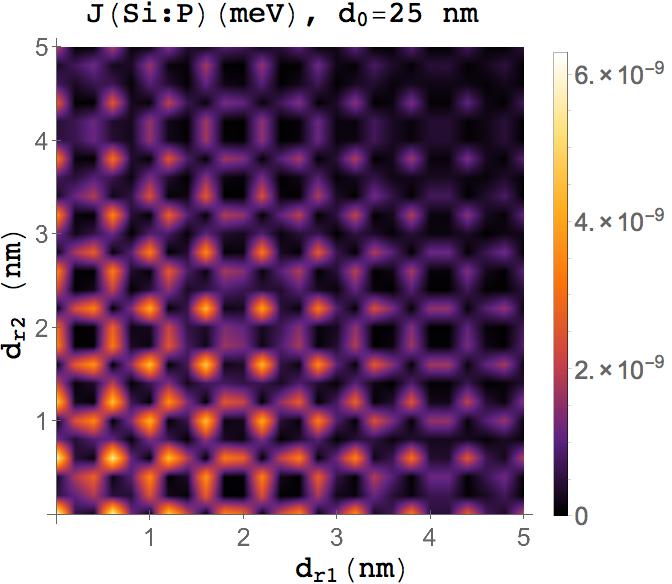} &  \includegraphics[width=.24\textwidth]{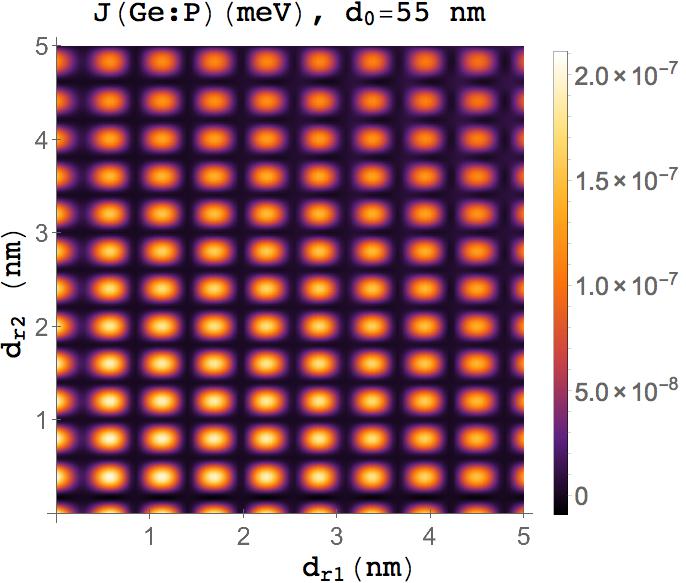}\\
\end{tabular}
\caption{Density plots of exchange couplings $J$ between neighboring donor spins as a function of misplacements $\textbf{d}_{r1}, \textbf{d}_{r2}$ in the plane orthogonal to the nominal donor separation $\textbf{d}_0$, i.e. $\textbf{d}_0 = [$20 $\text{nm},0,0]$ for Si:P (left) and $\textbf{d}_0 = [50/\sqrt{2}$ $ \text{nm},50/\sqrt{2}$ $ \text{nm},0]$ for Ge:P (right). $\{\textbf{d}_{r1}, \textbf{d}_{r2}\}$ are sampled with steps of 0.2~nm along each of the two directions; these calculated values are then interpolated for the purpose of illustration. Top panels refer to a situation of no parallel misplacement along $\textbf{d}_0$, bottom panels have a parallel misplacement of 5 nm. It is clear that the two-dimensional oscillatory pattern is preserved intact between bottom and top pictures, as explained in the text. The period of the oscillations is slightly larger in Si:P, as a result of the different conduction band structure. Remarkably, higher $J$ values survive for longer misplacement vectors in the plane for Ge:P than for Si:P.
}
\label{fig:patterns}
\end{figure}

As a result, in both of the optimal geometries just outlined for Si and Ge, the non-oscillating terms pertaining to valley-differences orthogonal to $\textbf{d}_0$ make the largest contributions to $J$. We are thus led to separate our analysis of parallel $\textbf{d}_r\parallel\textbf{d}_0$ and orthogonal $\textbf{d}_r\perp\textbf{d}_0$ misplacements from the nominal donor separation, since the qualitative dependence of $J$ is very different in the two cases. As the parallel misplacement is increased, the only effect is a monotonic decrease of the overall coupling due to the weaker overlap of the electronic densities. On the other hand for a fixed parallel displacement, $J$ oscillates markedly as a function of orthogonal misplacement. This is indeed clear from Fig.~\ref{fig:patterns}, that shows density plots of $J$ as a function of transverse misplacements for two different fixed parallel misplacements, for both Si:P and Ge:P donors. From our full numerical calculations, we find that to a good approximation it is possible to factor out the effects of parallel and transverse misplacements: $\log[J(\textbf{d}_r)/J(\textbf{d}_r=0)]\approx \log[e^{-1.82(\textbf{d}_r\cdot\hat{d}_0)/\bar{a}_l}]+\log f(\textbf{d}_r\wedge\hat{d}_0)$, where $f(x)$ is an highly oscillatory function of its argument.

Fig.~\ref{fig:patterns} also helps us to quantify how large the separations between Ge:P donors can be, while providing the same mean coupling of a Si:P donor ensemble. We can calculate two sets of $J$ couplings: the first corresponds to a statistical ensemble of Si:P donor pairs with nominal separation $d_0=20$ nm along the $[100]$ crystallographic direction but with random transverse placement error $\textbf{d}_r$ that is uniformly distributed in the square $[0, \{0,5 $ $\text{nm}\},\{0,5 $ $\text{nm}\}]$, the second to an ensemble of Ge:P donor pairs with nominal separation $d_0=50$ nm along the $[110]$ crystallographic direction and random transverse placement error $\textbf{d}_r$ that is uniformly distributed in the square of edge length 5 nm in the (110) plane with a vertex at the origin. For symmetry reasons, our analysis is also directly valid for the corresponding squares with edge length 10 nm centered at the origin. If we also include parallel misplacements, we get the  full distributions of $J$-couplings in Si:P and Ge:P for misplacements in the 5-nm cube, which are collected in the histograms in Fig.~\ref{fig:histos}. The mean coupling is computed to be $3.5\times10^{-9}$ meV in the Si:P cluster and $3\times10^{-9}$ meV in the Ge:P cluster: we reiterate that these similar mean values correspond to an average donor separation that increases from 20 nm in Si:P to 50 nm in Ge:P. This is compatible with the ratio of the transverse effective Bohr radii in the two materials $\bar{a}^{Ge:P}_l/\bar{a}^{Si:P}_l\approx 2.7$. From the fabrication point of view, this feature promises a huge relative improvement, as the more relaxed lengthscales in Ge would directly allow for much more space between donors for control gates and readout devices. 

\begin{figure}[h]
\centering
\includegraphics[width=.5\textwidth]{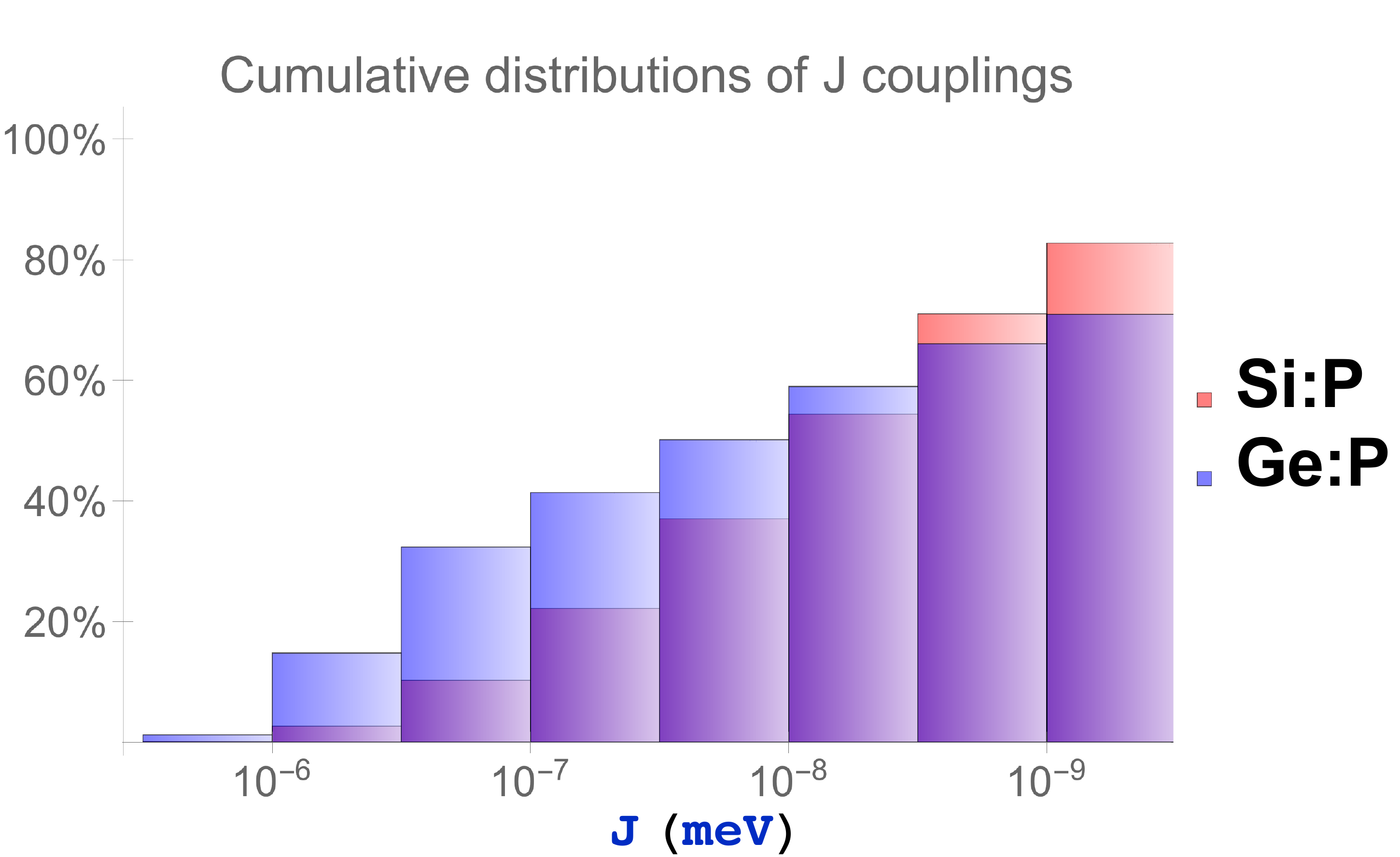}\\
\caption{Comparison of the cumulative distributions of the ensembles of Ge:P and Si:P couplings (on a log-scale) across the top four orders of magnitude of $J$. As explained in the text, the larger skew of the Ge:P distribution in Fig.~\ref{fig:histos} implies higher probabilities for Ge:P couplings to lie within the top two orders of magnitude. Larger couplings are desirable for faster two-qubit gates, and current schemes for uniform control of such gates in a cluster require that $J$ values vary at most within two orders of magnitude in the cluster. In this sense, Ge:P donors provide a significant advantage over Si:P donors.
}
\label{fig:cdf}
\end{figure}

Direct inspection of the histograms points out another interesting aspect of our comparison, related to the required precision of donor positioning in both semiconductor materials. The distribution of $J$ couplings of Ge:P donors is clearly more skewed to higher coupling values than in Si:P, with about a $33\%$ chance of a Ge:P pair yielding a coupling within one order of magnitude from the strongest in the set. This probability drops to less than $11\%$ for Si:P pairs. A more quantitative comparison is presented in Fig.~\ref{fig:cdf}, where cumulative distributions of the coupling ensembles from Si:P and Ge:P are shown on top of one another. After ion implantation, it is reasonable to assume that device characterization should identify the donors that are best suited for hosting quantum information: current schemes for exchange-based two-qubit gates\cite{twogatehyp,PhysRevB.93.035306} require that the range of $J$ couplings is at least contained within about two orders of magnitude, if the same kind of control is to be used across the device (or the cluster). Of course, larger couplings are more appealing as they imply the possibility of faster gates, more robust against single-qubit decoherence. Fig.~\ref{fig:cdf} shows that with the same coupling imprecision, pairs with a coupling within one order of magnitude of the maximum are significantly more likely to be realized in Ge:P than in Si:P. This advantage remains, albeit diminished, even if the chosen implementation architecture is robust to more than two orders of magnitude in $J$ (it is hard to imagine, however, how uniform control could be extended to clusters with more than two orders of magnitude variations of $J$). Thus, the requirements on precise donor positioning would be significantly less demanding for Ge:P than for Si:P donors. 

We finally remark that the considerations presented in this section do not change if other sets of donor distances are considered: as previously explained, if $\textbf{d}_0$ is decreased the main effect on $J$ is a simple monotonic exponential gain that is fit well with $J\sim e^{-1.82(\textbf{d}_0)/\bar{a}_l}$ (which can be compared to the analytical approximate formula in Ref.~\onlinecite{bhatt}).

\section{Conclusions}
Starting from a complete theory of shallow Ge-donors, based on an improved MV-EMT with full inclusion of the Bloch states' structure, we have assessed aspects of the performance of both single and two-qubit operations with donor spins in Ge. Close agreement with recent experiments in Ref.~\onlinecite{sigillito16} validates our predictions and insight into the underlying physical mechanisms. Maximum hyperfine coupling detunings in Ge are seen to closely match those in Si, but the giant tunability of the Zeeman splittings of Ge-donors suggests that nanosecond selective electrical spin manipulation could be achieved within realistic experimental settings. We related this feature to both the large conduction band anisotropy and the weak valley-orbit coupling in Ge. Further, the same average spin exchange coupling between two Si:P donors could be achieved with two Ge:P donors almost three times farther apart, which promises much more space for fabricating control gates and readout devices in heterostructures. Finally, in the optimal geometric configuration we show that the relative variation of the $J$ coupling across a realistic ensemble of randomly misplaced donors is smaller than in Si, which would make uniform control of large qubit clusters relatively easier.

With millisecond spin coherence times recently demonstrated by Ge-donor spins, and with fast developing spin transport in Ge heterostructures~\cite{getransport}, Ge is close to matching several requirements for quantum computing. We hope that our calculations of the high speed of the electrical qubit manipulations and the advantages in two-qubit coupling raise interest in further experimental investigation of donors in Ge as a promising alternative to the more established Si framework.

\begin{acknowledgements} We wish to thank Stephen Lyon and Anthony Sigillito for drawing our attention to the problem of describing donor states in Ge. This research was funded by the joint EPSRC (EP/I035536)/ NSF~(DMR-1107606) Materials World Network grant (GP, BWL), the EPSRC grant EP/K025562 (BWL). GP thanks the University of St Andrews and the EPSRC for a Doctoral Prize Fellowship.
\end{acknowledgements}

\section*{Appendix A: Valley structure of the donor spin exchange coupling}
It is convenient to write down the spin exchange splitting between two adjacent donor electrons, evaluted in the Heitler-London approximation~\cite{hl}, in a way that highlights most clearly its valley structure. The latter, in fact, determines predominantly the behaviour of $J$ as a function of the inter-donor separation. If $\Psi(\textbf{r}_{1}-\textbf{R}_{a})$ and $\Psi(\textbf{r}_{2}-\textbf{R}_{b})$ indicate the single-particle electron wave functions of electron 1 and 2, with $\textbf{r}_{i}$ defining the electronic coordinates, $\textbf{R}_{j}$ the fixed positions of the nuclear pair, and $\textbf{R}_{a}-\textbf{R}_{b}\equiv \textbf{d}$, we can write
\begin{equation}\label{spiegoj}
J=\frac{2}{1- {\cal S}^{4}}[{\cal S}^{2}(W + C_{D})-C_{I}+{\cal S}{\cal T}],
\end{equation}
where the single-particle feature $W$ is
\begin{equation}
W=-2 \langle \Psi_{a}|\frac{e^{2}}{\epsilon_{Si}|\textbf{r}-\textbf{R}_{b}|}|\Psi_{a}\rangle = - 2 \langle \Psi_{b}|\frac{e^{2}}{\epsilon_{Si}|\textbf{r}-\textbf{R}_{a}|}|\Psi_{b}\rangle, \end{equation}
the two-particle Coulombic integrals are
\begin{eqnarray}\nonumber
&C_{D}=\langle \Psi_{a}(\textbf{r}_{1})\Psi_{b}(\textbf{r}_{2})|\dfrac{e^{2}}{\epsilon_{Si}|\textbf{r}_{1}-\textbf{r}_{2}|}|\Psi_{a}(\textbf{r}_{1})\Psi_{b}(\textbf{r}_{2})\rangle,\\
&C_{I}=\langle \Psi_{a}(\textbf{r}_{1})\Psi_{b}(\textbf{r}_{2})|\dfrac{e^{2}}{\epsilon_{Si}|\textbf{r}_{1}-\textbf{r}_{2}|}|\Psi_{a}(\textbf{r}_{2})\Psi_{b}(\textbf{r}_{1})\rangle,
\end{eqnarray} 
and
\begin{equation}
{\cal T}=-2 \langle \Psi_{a}|\frac{e^{2}}{\epsilon_{Si}|\textbf{r}-\textbf{R}_{a}|}|\Psi_{b}\rangle=-2 \langle \Psi_{a}|\frac{e^{2}}{\epsilon_{Si}|\textbf{r}-\textbf{R}_{b}|}|\Psi_{b}\rangle.
\end{equation}
Now we discuss the qualitative behaviour of each of this quantities in turn. The overlap ${\cal S}=\langle \Psi_{a}|\Psi_{b}\rangle$ has been already discussed in the main text. $W$ does not oscillate significantly with the inter-donor separation: as a single-particle one-site integral, the interfering valley terms containing $F_{\mu}(\textbf{r}-\textbf{R}_{a})F_{\nu}(\textbf{r}-\textbf{R}_{a})$ with $\mu\neq\nu$ are much less important than the $\mu=\nu$ ones, which decrease monotonically with increasing $d$. $C_{D}$ represents the so-called direct exchange integral, which weights the electron-electron repulsion with the on-site donor densities: as such, it monotonically decreases as the electronic clouds overlap less. Unlike the valley structure pointed out for the overlap ${\cal S}$, the lack of inter-donor integrals prevents the appearance of $d$-oscillating terms, as 
\begin{align}\nonumber
\int \hspace{-2mm} d\textbf{r}_{1}d\textbf{r}_{2}\hspace{-1mm}\left(\hspace{-1mm}\sum_{\mu}\hspace{-1mm} u(\textbf{k}_{\mu},\textbf{r}_{1}-\textbf{R}_{a}) F_{\mu}(\textbf{r}_{1}-\textbf{R}_{a})\cos[\textbf{k}_{\mu}\hspace{-1mm}\cdot\hspace{-1mm} (\textbf{r}_{1}-\textbf{R}_{a})]\right)^{2} \\ \nonumber \times  \left(\sum_{\nu} u(\textbf{k}_{\nu},\textbf{r}_{2}-\textbf{R}_{b}) F_{\nu}(\textbf{r}_{2}-\textbf{R}_{b})\cos[\textbf{k}_{\nu}\cdot (\textbf{r}_{2}-\textbf{R}_{b})]\right)^{2}\approx \\ \dfrac{1}{4}\int d\textbf{r}_{1}d\textbf{r}_{2} \left(\sum_{\mu} F^{2}_{\mu}(\textbf{r}_{1}-\textbf{R}_{a})\right)\left(\sum_{\nu} F^{2}_{\nu}(\textbf{r}_{2}-\textbf{R}_{b})\right),\nonumber
\end{align}
where the cosines have been averaged over the integration region, as the lengthscales over which the envelopes vary significantly are much larger than $1/k_{0}$. Finally,
$C_{I}$ is the indirect exchange integral, which is fundamentally an inter-donor feature, and as such it reproduces the kind of valley pattern underlining the overlap ${\cal S}$, but `squared' for each of the electronic coordinates. After neglecting all spatial terms that oscillate strongly over the integration regions, we can write 
\begin{align}\label{excsimple}\nonumber
C_{I}\approx &\dfrac{1}{(\symbol{35} \text{valleys})^2}\int d\textbf{r}_{1}d\textbf{r}_{2}F_{\mu}(\textbf{r}_{1}-\textbf{R}_{a})F_{\nu}(\textbf{r}_{2}-\textbf{R}_{b})\\ \nonumber &F_{\nu}(\textbf{r}_{2}-\textbf{R}_{a}) \times F_{\mu}(\textbf{r}_{1}-\textbf{R}_{b})\frac{e^{2}}{|\textbf{r}_{1}-\textbf{r}_{2}|}e^{i(\textbf{k}_{\mu}-\textbf{k}_{\nu})\cdot \textbf{d}}.
\end{align}
Since time-reversal symmetry guarantees that $F_{\mu}=F_{-\mu}$, we can restate this last result as 
\begin{equation}\label{excfinal}
C_{I}\approx \dfrac{1}{(\symbol{35} \text{valleys})^2}C^{\mu\nu}_{I}(\textbf{d})\cos(\textbf{k}_{\mu}-\textbf{k}_{\nu})\cdot \textbf{d}, \end{equation} with
\begin{align}
\nonumber C^{\mu\nu}_{I}(\textbf{d})\equiv & \int d\textbf{r}_{1}d\textbf{r}_{2}F_{\mu}(\textbf{r}_{1}-\textbf{R}_{a})F_{\nu}(\textbf{r}_{2}-\textbf{R}_{b})F_{\nu}(\textbf{r}_{2}-\textbf{R}_{a})\\ &\times F_{\mu}(\textbf{r}_{1}-\textbf{R}_{b})\dfrac{e^{2}}{|\textbf{r}_{1}-\textbf{r}_{2}|}.
\end{align}
The inter-donor matrix element ${\cal T}$ displays essentially the same valley structure as ${\cal S}$, as the Fourier transform of the Coulomb potential is not able to couple significantly different valleys $\mu\neq\nu$.
\vspace{3.7cm}




\end{document}